\documentclass[amssymb,aps,eqsecnum]{revtex4}
\voffset= + 0.97in
\hoffset= + 0.15in
\usepackage{amsmath}
\usepackage{enumerate}
\usepackage{graphicx}
\DeclareGraphicsExtensions{.pdf,.png,.jpg}
\begin{document}
\def \brho {{\rho} \hskip -4.5pt { \rho}}
\def \bnab {{\bf\nabla}\hskip-8.8pt{\bf\nabla}\hskip-9.1pt{\bf\nabla}}
\title{Continuum electrodynamics and the Abraham--Minkowski momentum controversy}
\vskip -0.4cm
\author{Michael E. Crenshaw}
\thanks{michael.e.crenshaw4.civ@mail.mil}
\maketitle
\vskip -0.9cm
\centerline{US Army Aviation and Missile Research, Development, and Engineering Center}
\centerline{RMDR-WDS-W, Redstone Arsenal, AL 35898, USA}
\vskip 0.5cm
\centerline{[Published in Optical Trapping and Optical Micromanipulation XII, Proc. SPIE 9548, 95480J (25 Aug 2015)]}
\vskip 0.5cm
\centerline{\bf ABSTRACT}
\vskip 0.3cm
\par
Continuum electrodynamics is an axiomatic formal theory based
on the macroscopic Maxwell equations and the constitutive relations.
We apply the formal theory to a thermodynamically closed system
consisting of an antireflection coated block of dielectric
situated in free-space and illuminated by a quasimonochromatic field.
We show that valid theorems of the formal theory are proven false by
relativity and by conservation laws.
Then the axioms of the formal theory are proven false at a fundamental
level of mathematical logic.
We derive a new formal theory of continuum electrodynamics for
macroscopic electric and magnetic fields in a four-dimensional flat
non-Minkowski material spacetime in which the speed of light is $c/n$.
\par
\section{Macroscopic Maxwell's equations and relativity}
\par
By the middle of the 19th century, the results of more than two
centuries of electric and magnetic experiments had been codified into a
set of equations bearing the names of Gauss, Amp\`ere, Faraday, and,
occasionally, Thompson.
In 1864, Maxwell postulated a correction to the Amp\`ere equation making
it consistent with time-dependent fields and leading to the prediction
of electromagnetic wave phenomena.
In honor of his achievement, the empirical equations of electrodynamics
are known collectively as the Maxwell equations and are regarded as
physical laws.
Maxwell's equations appear in a microscopic form that applies to
electric and magnetic fields in the vacuum and appear, separately, in a
macroscopic form as the equations of motion for macroscopic electric and
magnetic fields in continuous media.
Then the macroscopic Maxwell equations, along with the constitutive 
relations, are the axioms of theoretical continuum electrodynamics.
New theorems are derived by applying the ordinary operations of algebra,
linear algebra, and calculus to the axioms and subsequent theorems.
The axioms and the rules for combining axioms and theorems to form new
theorems constitute an axiomatic formal theory of continuum
electrodynamics.
\par
The axiomatic formal theory of continuum electrodynamics is based
on the macroscopic Maxwell (Maxwell--Heaviside--Minkowski) equations
\begin{subequations}
\begin{equation}
\nabla\times{\bf H}-\frac{1}{c}\frac{\partial{\bf D}}{\partial t}=0
\label{EQpp1.01a}
\end{equation}
\begin{equation}
\nabla\cdot{\bf B}=0
\label{EQpp1.01b}
\end{equation}
\begin{equation}
\nabla\times{\bf E}+\frac{1}{c}\frac{\partial{\bf B}}{\partial t}=0
\label{EQpp1.01c}
\end{equation}
\begin{equation}
\nabla\cdot{\bf D}=0
\label{EQpp1.01d}
\end{equation}
\label{EQpp1.01}
\end{subequations}
and the linear constitutive relations
\begin{subequations}
\begin{equation}
{\bf D}= \varepsilon {\bf E}
\label{EQpp1.02a}
\end{equation}
\begin{equation}
{\bf B}= \mu {\bf H} \,.
\label{EQpp1.02b}
\end{equation}
\label{EQpp1.02}
\end{subequations}
We treat the case of a thermodynamically closed system consisting of a 
quasimonochromatic optical pulse incident on a simple linear dielectric.
The otherwise homogeneous dielectric is draped with a gradient-index 
antireflection coating.
The frequency of the incident radiation is required to be sufficiently
far from resonance that absorption is negligible and dispersion is
treated parametrically.
Then the macroscopic refractive index, $n$, is a time-independent real
scalar function of position.
The constitutive relations become
\begin{subequations}
\begin{equation}
{\bf D}= n^2 {\bf E}
\label{EQpp1.03a}
\end{equation}
\begin{equation}
{\bf B}= {\bf H}
\label{EQpp1.03b}
\end{equation}
\label{EQpp1.03}
\end{subequations}
by setting $\mu=1$ and $\varepsilon=(n({\bf r}))^2$ in
Eqs~(\ref{EQpp1.02}).
These simple material properties can be regarded as additional axioms in
our formal theory.
As a matter of clarity and concision, we are only treating simple linear
dielectrics.
Then, we can treat Eqs.~(\ref{EQpp1.03}) as replacements for the axioms
given by Eqs.~(\ref{EQpp1.02}).
The constitutive relations, Eqs.~(\ref{EQpp1.03}), are used to
substitutionally eliminate ${\bf H}$ and ${\bf D}$
from Eqs.~(\ref{EQpp1.01}).
The macroscopic Maxwell equations are
\begin{subequations}
\begin{equation}
\nabla\times{\bf B}-\frac{n^2}{c}\frac{\partial{\bf E}}{\partial t}=0
\label{EQpp1.04a}
\end{equation}
\begin{equation}
\nabla\cdot{\bf B}=0
\label{EQpp1.04b}
\end{equation}
\begin{equation}
\nabla\times{\bf E}+\frac{1}{c}\frac{\partial{\bf B}}{\partial t}=0
\label{EQpp1.04c}
\end{equation}
\begin{equation}
\nabla\cdot (n^2{\bf E}) =0 
\label{EQpp1.04d}
\end{equation}
\label{EQpp1.04}
\end{subequations}
for a simple linear dielectric medium.
After making the indicated substitutions, we commute the
time-independent refractive index with the temporal derivative in the
Maxwell--Amp\`ere law, Eq.~(\ref{EQpp1.04a}), and multiply the Faraday
law, Eq.~(\ref{EQpp1.04c}), by $n$.
Applying simple vector identities to commute curl and divergence
operators with a scalar function of position, we obtain
\begin{subequations}
\begin{equation}
\nabla\times{\bf B}-\frac{n}{c}\frac{\partial n{\bf E}}{\partial t}=0
\label{EQpp1.05a}
\end{equation}
\begin{equation}
\nabla\cdot{\bf B}=0
\label{EQpp1.05b}
\end{equation}
\begin{equation}
\nabla\times n{\bf E}+\frac{n}{c}\frac{\partial{\bf B}}{\partial t}=
\frac{\nabla n}{n} \times n{\bf E}
\label{EQpp1.05c}
\end{equation}
\begin{equation}
\nabla\cdot n{\bf E}= -\frac{\nabla n}{n}\cdot n{\bf E} \, .
\label{EQpp1.05d}
\end{equation}
\label{EQpp1.05}
\end{subequations}
Note that Eqs.~(\ref{EQpp1.05}) are valid theorems of the axiomatic
formal theory of continuum electrodynamics.
There are no assumptions or hidden axioms that enter the derivation
of Eqs.~(\ref{EQpp1.05}) from the stated axioms, Eqs.~(\ref{EQpp1.01})
with the constitutive relations, Eqs.~(\ref{EQpp1.03}).
\par
There appears to be nothing remarkable about Eqs.~(\ref{EQpp1.05})
because they are clearly equivalent to Eqs.~(\ref{EQpp1.01}) with
(\ref{EQpp1.03}).
On the other hand, it has been suggested \cite{BIprivate} that
Eqs.~(\ref{EQpp1.05}) violate the theory of special relativity.
The Lorentz factor that is associated with Minkowski spacetime is
\begin{equation}
\gamma_M=\frac{1}{\sqrt{1-\frac{v^2}{c^2}}} \, .
\label{EQpp1.06}
\end{equation}
This is true for fields in a dielectric medium \cite{BIMoller},
as well as for fields in the vacuum, and is validated by the 
Fizeau experiment \cite{BIFiz}.
Meanwhile, the material Lorentz factor \cite{BIRosen,BIFinn},
\begin{equation}
\gamma_d=\frac{1}{\sqrt{1-\frac{n ^2v^2}{c^2}}} \, ,
\label{EQpp1.07}
\end{equation}
is associated with Eqs.~(\ref{EQpp1.05}) and can be considered to be 
incorrect based on the correct Lorentz factor, Eq.~(\ref{EQpp1.06}).
Now, the two sets of equations, Eqs.~(\ref{EQpp1.05}) and
Eqs.~(\ref{EQpp1.01}) with (\ref{EQpp1.03}), are equivalent within the
current theory of continuum electrodynamics and are equally correct.
Because the two sets of equations have different relativistic
properties, at least one set of equations is wrong.
Then both sets of equations are wrong by equivalence.
Consequently, both the valid theorem of continuum electrodynamics,
Eqs.~(\ref{EQpp1.05}), and the axioms of continuum electrodynamics,
Eqs.~(\ref{EQpp1.01}) and (\ref{EQpp1.03}), are false.
This is a mathematical fact.
\par
\section{The Abraham--Minkowski controversy}
\par
In 1908, Minkowski \cite{BIMin} proposed an energy--momentum tensor for
electromagnetic fields in a linear medium.
The following year, Abraham \cite{BIAbr} argued from symmetry to propose
a correction to the momentum density that appears in the Minkowski
energy--momentum tensor.
The ``corrected'' tensor is known as the Abraham energy--momentum
tensor.
The consequent Abraham--Minkowski controversy refers to the
century-long inability of scientists to conclusively identify the role
of electromagnetic momentum for the propagation of light in linear
media \cite{BIAMC2,BIAMC2x,BIAMC3,BIAMC4,BIPfeif}.
To be sure, the total momentum \cite{BIPfeif,BIGord,BICrenxxx},
\begin{equation}
{\bf G}_T=\int_{\sigma} \frac{ n{\bf E}\times{\bf B}}{c} dv ,
\label{EQpp2.01}
\end{equation}
for a quasimonochromatic pulse incident on a rare or
antireflection-coated transparent dielectric has never been in doubt
because the momentum density integrated over a region that is extended
to all-space ${\sigma}$ is conserved in a closed system.
Instead, the disputed issue is the amount of the total momentum that
remains with the propagating electromagnetic field and the amount of
total momentum that is transferred to the matter by various forms of
radiation pressure.
\par
The Abraham--Minkowski momentum controversy is very much a part of 
axiomatic continuum electrodynamics.
We want to revisit the electromagnetic continuity equations from the
point of view of the axiomatic formal theory presented in Section I.
Subtracting the scalar product of Eq.~(\ref{EQpp1.04a}) with ${\bf E}$
from the scalar product of Eq.~(\ref{EQpp1.04c}) with ${\bf B}$, one
obtains 
\begin{equation}
\frac{1}{c}\frac{\partial }{\partial t} \left [
\frac{1}{2}\left ( n^2{\bf E}^2+{\bf B}^2\right )
 \right ]+\nabla\cdot c\frac{{\bf E}\times{\bf B}}{c}= 0 \, .
\label{EQpp2.02}
\end{equation}
Known as Poynting's theorem, Eq.~(\ref{EQpp2.02}) has the appearance of
being an energy continuity equation.
Similarly, we can start with the vector product of $n^2{\bf E}$ with
Eq.~(\ref{EQpp1.04c}) and subtract the vector product of
Eq.~(\ref{EQpp1.04a}) with ${\bf B}$ to obtain 
\begin{equation}
\frac{1}{c}\frac{\partial( n^2{\bf E}\times{\bf B})}{\partial t}+
\nabla\cdot{\bf W}= {\bf f}_M \, ,
\label{EQpp2.03}
\end{equation}
which has the appearance of a momentum continuity equation.
Here,
\begin{equation}
W_{ij}=-n^2 E_i E_j-B_iB_j+
\frac{1}{2} (n^2{\bf E}^2+{\bf B}^2)\delta_{ij}
\label{EQpp2.04}
\end{equation}
is the Maxwell stress tensor and
\begin{equation}
{\bf f}_M=-\frac{n^2{\bf E}^2}{2} \frac{\nabla (n^2)}{n^2}
\label{EQpp2.05}
\end{equation}
is a force density.
Both Eq.~(\ref{EQpp2.02}) and Eq.~(\ref{EQpp2.03}) are valid theorems of
continuum electrodynamics.
As a matter of linear algebra, the scalar equation,
Eq.~(\ref{EQpp2.02}), and the three orthogonal components of the vector
equation, Eq.~(\ref{EQpp2.03}), can be written as a single matrix
differential equation
\begin{equation}
\partial_{\beta} T_M^{\alpha\beta} =f^{\alpha}_M \, .
\label{EQpp2.06}
\end{equation}
In Eq.~(\ref{EQpp2.06}), 
\begin{equation}
\partial_{\beta}=\left ( \frac{1}{c}\frac{\partial}{\partial t},
\frac{\partial}{\partial x},\frac{\partial}{\partial y},
\frac{\partial}{\partial z} \right ) 
\label{EQpp2.07}
\end{equation}
is an operator,
\begin{equation}
T_M^{\alpha\beta}=
\left [
\begin{matrix}
(n^2{\bf E}^2+{\bf B}^2)/2 &({\bf E}\times{\bf B})_1
&({\bf E}\times{\bf B})_2 &({\bf E}\times{\bf B})_3
\cr
(n^2{\bf E}\times{\bf B})_1   &W_{11}     &W_{12}      &W_{13}
\cr
(n^2{\bf E}\times{\bf B})_2    &W_{21}      &W_{22}      &W_{23}
\cr
(n^2{\bf E}\times{\bf B})_3    &W_{31}      &W_{32}      &W_{33}
\cr
\end{matrix}
\right ] 
\label{EQpp2.08}
\end{equation}
is a matrix, and
\begin{equation}
f^{\beta}_M=\left ( 0,-\frac{(n{\bf E})^2}{2}\frac{\nabla (n^2)}{n^2}
\right ) ^T 
\label{EQpp2.09}
\end{equation}
is a matrix.
\par
The matrix, Eq.~(\ref{EQpp2.08}), which has the appearance of being an 
energy--momentum tensor is known as the Minkowski energy--momentum
tensor.
In a thermodynamically closed system, the energy--momentum tensor is
required to be diagonally symmetric so that angular momentum is 
conserved \cite{BICT,BIJack}.
The Minkowski energy--momentum tensor is not symmetric and  Abraham 
proposed a diagonally symmetric matrix
\begin{equation}
T_A^{\alpha\beta}=
\left [
\begin{matrix}
(n^2{\bf E}^2+{\bf B}^2)/2 &({\bf E}\times{\bf B})_1
&({\bf E}\times{\bf B})_2 &({\bf E}\times{\bf B})_3
\cr
({\bf E}\times{\bf B})_1   &W_{11}     &W_{12}      &W_{13}
\cr
({\bf E}\times{\bf B})_2    &W_{21}      &W_{22}      &W_{23}
\cr
({\bf E}\times{\bf B})_3    &W_{31}      &W_{32}      &W_{33}
\cr
\end{matrix}
\right ] 
\label{EQpp2.10}
\end{equation}
for the energy--momentum tensor.
Although physically motivated by conservation laws, Abraham's conjecture
creates other conservation issues with the formal theory of continuum 
electrodynamics. 
\par
The modern consensus is that the system represented by the macroscopic
Maxwell theory is not thermodynamically closed, but is, in fact, an
electromagnetic subsystem that is coupled to a material subsystem.
In this scenario, the Minkowski force density $f^{\alpha}_M$ transfers
momentum from the electromagnetic subsystem, $T$, to a material
subsystem, $R$.
Then,
\begin{equation}
\partial_{\beta} R^{\alpha\beta} =-f^{\alpha}_M 
\label{EQpp2.11}
\end{equation}
such that the total field plus material system thermodynamically closed,
\begin{equation}
\partial_{\beta}\left ( T_M^{\alpha\beta}+R^{\alpha\beta} \right )=0\, .
\label{EQpp2.12}
\end{equation}
Then, motivated by the conservation laws, one assumes some properties
for the material subsystem $R$.
In Pfeifer, et al. \cite{BIPfeif}, for example, 
\begin{equation}
R^{\alpha\beta}=
\left [
\begin{matrix}
c^2 \rho   &c \rho v_1    &c \rho v_2   &c \rho v_3
\cr
c \rho v_1 +(1-n^2)({\bf E}\times{\bf B})_1  & \rho v_1 v_1 
& \rho v_1 v_2  & \rho v_1 v_3
\cr
c \rho v_2 +(1-n^2)({\bf E}\times{\bf B})_2  & \rho v_2 v_1 
& \rho v_2 v_2  & \rho v_2 v_3
\cr
c \rho v_3 +(1-n^2)({\bf E}\times{\bf B})_3  & \rho v_3 v_1 
& \rho v_3 v_2  & \rho v_3 v_3
\cr
\end{matrix}
\right ] \, ,
\label{EQpp2.13}
\end{equation}
where $\rho$ is the mass density and ${\bf v}$ is the velocity field.
In order for the total momentum, Eq.~(\ref{EQpp2.01}), to be conserved,
we must assume the material momentum density to be \cite{BIPfeif}
\begin{equation}
\rho{\bf v}=(n-1) \frac {{\bf E}\times{\bf B}}{c} \,.
\label{EQpp2.14}
\end{equation}
Now, $f^{\alpha}_M$ can be made vanishingly small for an arbitrarily
large homogeneous medium draped with a gradient-index antireflection
coating.
In that limit, the electromagnetic and material subsystems are not
coupled and, consequently, 
\begin{equation}
\partial_{\beta} T_M^{\alpha\beta}=0
\label{EQpp2.15}
\end{equation}
\begin{equation}
\partial_{\beta} R^{\alpha\beta}=0 \, .
\label{EQpp2.16}
\end{equation}
Then the material tensor
\begin{equation}
R^{\alpha\beta}=
\left [
\begin{matrix}
c^2 \rho   &c \rho v_1    &c \rho v_2   &c \rho v_3
\cr
c \rho v_1 & \rho v_1 v_1 & \rho v_1 v_2  & \rho v_1 v_3
\cr
c \rho v_2 & \rho v_2 v_1 & \rho v_2 v_2  & \rho v_2 v_3
\cr
c \rho v_3 & \rho v_3 v_1 & \rho v_3 v_2  & \rho v_3 v_3
\cr
\end{matrix}
\right ]
\label{EQpp2.17}
\end{equation}
is the dust tensor that is the total energy--momentum tensor for the
unimpeded (force-free) flow of non-interacting material particles.
Furthermore, the vanishingly small force density means that there is no
appreciable source or sink of momentum.
Consequently, the Minkowski tensor is the total energy--momentum tensor
of a thermodynamically closed system and that means that the Minkowski
momentum 
\begin{equation}
{\bf G}_M=\int_{\sigma} \frac{ n^2{\bf E}\times{\bf B}}{c} dv 
\label{EQpp2.18}
\end{equation}
is conserved.
Although the mathematics, $\partial_{\beta} T_M^{i\beta} =0$,
tell us that the Minkowski momentum is
conserved, it is well-known that the Minkowski momentum is a factor of
$n$ greater than the incident momentum \cite{BIPfeif,BIGord,BICrenxxx}.
Then the continuity equation, Eq.~(\ref{EQpp2.06}), is false in the
limit of an arbitrarily large homogenous medium draped with a
gradient-index antireflection coating.
Because a valid theorem of continuum electrodynamics,
Eqs.~(\ref{EQpp2.06}), has been proven false, the axioms of the theory,
the macroscopic Maxwell equations, are also false. 
In addition, the absence of diagonal symmetery and the consequent lack
of angular momentum conservation also disproves the macroscopic Maxwell
equations by disproving a valid theorem that has been derived from these
axioms.
\par
The entire history of the Abraham--Minkowski controversy is the search
for valid conservation laws for energy and momentum.
Recasting the debate in the language of axioms and formal logic imposes
the discipline that is necessary in order to prevent the introduction of
hypothetical forces and momentums.
Instead, the Maxwellian theory is formally proven to be invalid because
it leads to demonstrably incorrect results.
The search for valid conservation laws for energy and momentum must be
taken outside the Maxwellian, and equivalent, formulations of continuum
electrodynamics.
\par
\section{What to do?}
\par
The phenomenological Maxwell--Heaviside--Minkowski equations have 
carried us through a century-and-a-half of remarkable progress in
science.
However, we have demonstrated with examples from relativity and
energy/momentum conservation that there are situations in which the
axiomatic formal theory of continuum electrodynamics is contradicted.
While we can, and do, work around these problems, we would seek a theory
of continuum electrodynamics with innate conformity to other fundamental
physical principles.
Despite almost certain objections, there is now clear and compelling
evidence of the need to a construct a completely new theory of
continuum electrodynamics.
Ideally, the best place to start is with one of the other fundamental
physical principles, like conservation, relativity, or field theory.
While these principles are intrinsic to the vacuum, the implementations
of these fundamental principles in a dielectric environment lead to
dogmatic arguments about the momentum of electromagnetic fields and a
perpetuation of the Abraham--Minkowski momentum controversy.
Then we must jettison the phenomenological Maxwell equations, the
constitutive relations, and the variations of the macroscopic Maxwell
equations.
It is necessary to derive the equations of motion for macroscopic
fields and to re-derive the other fundamental physical principles 
for fields in a region of space in which the speed of light is $c/n$.
\par
We define an inertial reference frame $S(x,y,z)$ that is located in 
free-space.
If a light pulse is emitted from the origin at time $t=0$, then
wavefronts are described by a mathematical formula
\begin{equation}
x^2+y^2+y^2-\left ( ct \right )^2 =0.
\label{EQpp3.01}
\end{equation}
The four-vector $(x_0,x,y,z)$ represents the position of a
point in a four-dimensional Minkowski vacuum spacetime
in which $x_0=ct$ is a timelike coordinate.
Dielectrics are mostly vacuum, consisting of tiny bits of matter
separated by relatively large distances.
Very small particles, like electrons, can travel through the dielectric
at speeds greater than $c/n$ because they mostly pass through the
interstitial space.
Likewise, the instantaneous speed of light in a dielectric is
$c$, although the effective speed of light is $c/n$ due to scattering.
In continuum electrodynamics, however, a dielectric is continuous at all
length scales.
In the rest frame of the dielectric, light travels at a uniform speed of
$c/n$ and wavefronts from a point light source are described by 
\begin{equation}
x^2+y^2+y^2-\left ( \frac{ct}{n} \right )^2 =0.
\label{EQpp3.02}
\end{equation}
The four-vector $(\bar x_0,x,y,z)$ represents the position of a
point in a flat four-dimensional non-Minkowski material spacetime
in which $\bar x_0=ct/n$ is a timelike coordinate.
Although the existence of a non-Minkowski material spacetime might be
viewed skeptically, Eq.~(\ref{EQpp3.02}), which defines the spacetime,
is solid ground.
\par
If we adopt Minkowski spacetime, the Lagrange equation for fields in a
linear medium is \cite{BICT,BIHillMlod}
\begin{equation}
\frac{d}{d x_0}\frac{\partial{\cal L}}
{\partial (\partial A_j /\partial x_0)}
=\frac{\partial {\cal L}}{\partial A_j}
-\sum_i\partial_{i}
\frac{\partial{\cal L}}{\partial(\partial_{i} A_j )} \, .
\label{EQpp3.03}
\end{equation}
Applying Eq.~(\ref{EQpp3.03}) to the Lagrangian density of the
electromagnetic field in the medium results in the usual macroscopic
Maxwell equations \cite{BICT}.
\par
As discussed above, a dielectric environment is distinct from a vacuum
environment.
Adopting the non-Minkowski material spacetime, we derive \cite{BIPre}
\begin{equation}
\frac{d}{d \bar x_0}\frac{\partial{\cal L}}
{\partial (\partial A_j /\partial \bar x_0)}
=\frac{\partial {\cal L}}{\partial A_j}
-\sum_i\partial_{i}
\frac{\partial{\cal L}}{\partial(\partial_{i} A_j )}
\label{EQpp3.04}
\end{equation}
as the Lagrange equation for electromagnetic fields in a linear medium.
Applying Eq.~(\ref{EQpp3.04}) to the Lagrangian density of the
electromagnetic field in the medium 
\begin{equation}
{\cal L}=
\frac{1}{2}
\left ( \left (
\frac{\partial{\bf A}}{\partial \bar x_0} \right )^2
-(\nabla\times{\bf A})^2 \right ) \, ,
\label{EQpp3.05}
\end{equation}
one defines the canonical momentum density ${\bf \Pi}$ in terms of
components as
\begin{equation}
\Pi_j=\frac{\partial{\cal L}}{\partial (\partial A_j/\partial\bar x_0)}
=\frac{\partial A_j}{\partial \bar x_0} 
\label{EQpp3.06}
\end{equation}
such that
\begin{equation}
{\bf \Pi}= \frac{\partial {\bf A}}{\partial \bar x_0} \, .
\label{EQpp3.07}
\end{equation}
Then the Lagrange equation takes the familiar form of the wave equation
\begin{equation}
\nabla\times(\nabla\times{\bf A})+
\frac{\partial^2{\bf A}}{\partial\bar x_0^2}=0 \, .
\label{EQpp3.08}
\end{equation}
We define a macroscopic magnetic field
\begin{equation}
{\bf B}=\nabla\times{\bf A} \, .
\label{EQpp3.09}
\end{equation}
Substituting the macroscopic fields ${\bf B}$ and ${\bf \Pi}$ into the
wave equation, Eq.~(\ref{EQpp3.08}), and performing vector calculus
operations on the macroscopic fields, Eqs.~(\ref{EQpp3.07}) and
(\ref{EQpp3.09}), and the wave equation, we obtain
\begin{subequations}
\begin{equation}
\nabla\times{\bf B}+\frac{\partial {\bf \Pi}}{\partial \bar x_0}=0
\label{EQpp3.10a}
\end{equation}
\begin{equation}
\nabla\cdot{\bf B}=0
\label{EQpp3.10b}
\end{equation}
\begin{equation}
\nabla\times {\bf \Pi}-\frac{\partial{\bf B}}{\partial \bar x_0}=
\frac{\nabla n}{n} \times {\bf \Pi}
\label{EQpp3.10c}
\end{equation}
\begin{equation}
\nabla\cdot {\bf \Pi}= \frac{\nabla n}{n}\cdot {\bf \Pi} \, .
\label{EQpp3.10d}
\end{equation}
\label{EQpp3.10}
\end{subequations}
as equations of motion for the macroscopic fields.
It might appear that these equations are equal to the phenomenological
Maxwell equations under a simple re-parameterization.
However, this is where we need to be careful not to be dragged back into
the discredited theory.
Specifically, classical continuum electrodynamics admits improper tensor
transformations of coordinates that alter the relativistic and
conservation properties of the macroscopic Maxwell equations.
We have abandoned classical continuum electrodynamic theory for that
reason and any transformations of Eqs.~(\ref{EQpp3.10}) are required to
maintain the tensor properties.
In particular, the timelike coordinate $\bar x_0=ct/n$ must remain
intact in order to preserve the spacetime.
The derivation from field theory gives Eqs.~(\ref{EQpp3.10}) their
privileged status and gives them physical relevance as equations of
motion for macroscopic electromagnetic fields in a flat non-Minkowski
material spacetime.
The suggested re-parameterization is forbidden.
\par
\section{Resolution of the Abraham--Minkowski controversy}
\par
The equations of motion for macroscopic fields in a continuous
dielectric medium, Eqs.~(\ref{EQpp3.10}), are theorems of the field
theory in a non-Minkowski material spacetime in which light travels at
speed $c/n$.
Now that the field theory has done its job and we have these theorems,
we can designate them as the axioms of our new theory of continuum
electrodynamics, keeping the field theory in the background.
The equations of motion for the macroscopic fields,
Eqs.~(\ref{EQpp3.10}), can be combined in the usual manner, using
algebra and calculus, to write electromagnetic energy and momentum
continuity equations
\begin{equation}
\frac{\partial }{\partial \bar x_0} \left [
\frac{1}{2}\left ( {\bf \Pi}^2+{\bf B}^2\right )
 \right ]+
\nabla\cdot \left ( {\bf B}\times{\bf \Pi}\right )
=\frac{\nabla n}{n} \cdot({\bf B}\times{\bf \Pi}) 
\label{EQpp4.01}
\end{equation}
\begin{equation}
\frac{\partial( {\bf B}\times{\bf \Pi})}{\partial \bar x_0}+
\nabla\cdot{\bf W}=
-{\bf \Pi}\left ({\bf \Pi}\cdot\frac{\nabla n}{n}\right ) \, ,
\label{EQpp4.02}
\end{equation}
where
\begin{equation}
W_{ij}=-\Pi_i\Pi_j-B_iB_j+
\frac{1}{2} ({\bf \Pi}^2+{\bf B}^2)\delta_{ij}
\label{EQpp4.03}
\end{equation}
is the Maxwell stress-tensor.
Then the continuity equations, Eqs.~(\ref{EQpp4.01}) and
(\ref{EQpp4.02}), can be written as
\begin{equation}
\bar\partial_{\alpha}T^{\alpha\beta}= f^{\alpha}
\label{EQpp4.04}
\end{equation}
where
\begin{equation}
T^{\alpha\beta}=
\left [
\begin{matrix}
({\bf \Pi}^2+{\bf B}^2)/2 &({\bf B}\times{\bf \Pi})_1
&({\bf B}\times{\bf \Pi})_2 &({\bf B}\times{\bf \Pi})_3
\cr
({\bf B}\times{\bf \Pi})_1   &W_{11}     &W_{12}      &W_{13}
\cr
({\bf B}\times{\bf \Pi})_2    &W_{21}      &W_{22}      &W_{23}
\cr
({\bf B}\times{\bf \Pi})_3    &W_{31}      &W_{32}      &W_{33}
\cr
\end{matrix}
\right ]
\label{EQpp4.05}
\end{equation}
is the energy--momentum tensor,
\begin{equation}
\bar\partial_{\beta}=
\left ( \frac{\partial}{\partial \bar x_0},
\frac{\partial}{\partial x},\frac{\partial}{\partial y},
\frac{\partial}{\partial z} \right )
\label{EQpp4.06}
\end{equation}
is the material four-divergence operator \cite{BIFinn}, and
\begin{equation}
f^{\alpha}=
\left (\frac{\nabla n}{n}\cdot({\bf B}\times{\bf \Pi}),
-{\bf \Pi}\left ({\bf \Pi}\cdot\frac{\nabla n}{n}\right ) 
\right ) 
\label{EQpp4.07}
\end{equation}
is the four-force density.
\par
The continuity equations, Eqs.~(\ref{EQpp4.01}), (\ref{EQpp4.02}), and
(\ref{EQpp4.04}), are valid theorems of a continuum electrodynamics
whose axioms, Eqs.~(\ref{EQpp3.10}), are themselves theorems of field
theory in a flat non-Minkowski material spacetime.
We can identify two limiting cases of particular interest.
In the limit of piecewise-homogeneous media, the field passes from 
one simple linear dielectric (or the vacuum) into a second such medium
through a planar interface at normal incidence where the field is
connected by boundary conditions at the interfaces between homogeneous
linear materials.
There is radiation pressure on the surface due to Fresnel reflection.
Here we consider the case of an arbitrarily large homogeneous simple
linear dielectric draped with a gradient index antireflection coating.
The gradient of the index of refraction $\nabla n$ can be made
sufficiently small that reflections can be neglected \cite{BICrenxxx}.
It is in this limit of an unimpeded continuous flow of electromagnetic
radiation in which there are no sources or sinks of energy or momentum
that the energy and momentum densities can be expressed through
continuity equations and a total energy--momentum tensor for a
thermodynamically closed system.
(Lorentz dipole forces belong to a microscopic model of non-interacting
particles, not to the macroscopic dielectric considered here and
certainly not to the axiomatic formalism of continuum electrodynamics.)
Taking the force density to be negligible, we obtain an energy
continuity equation and a momentum continuity equation
\begin{equation}
\frac{\partial }{\partial \bar x_0} \left [
\frac{1}{2}\left ( {\bf \Pi}^2+{\bf B}^2\right )
 \right ]+
\nabla\cdot \left ( {\bf B}\times{\bf \Pi}\right ) =0
\label{EQpp4.08}
\end{equation}
\begin{equation}
\frac{\partial( {\bf B}\times{\bf \Pi})}{\partial \bar x_0}+
\nabla\cdot{\bf W}=0
\label{EQpp4.09}
\end{equation}
expressing conservation of the total energy
\begin{equation}
U=\int_{\sigma} \frac{1}{2}\left ( {\bf \Pi}^2+{\bf B}^2\right ) dv
\label{EQpp4.10}
\end{equation}
and the total momentum
\begin{equation}
{\bf G}=\int_{\sigma} \frac{ {\bf B}\times{\bf \Pi} }{c} dv
\label{EQpp4.11}
\end{equation}
in an unimpeded (force-free) continuous flow of light in a
non-Minkowski material spacetime.
It is easy to verify that these conserved quantities are indeed
invariant in time, as is required.
Likewise, in the limit that the surface four-force density can be
neglected there is no coupling of energy or momentum to any material
subsystem.
Then, $T^{\alpha\beta}$, as shown in Eq.~(\ref{EQpp4.05}), is the
total energy--momentum tensor of a thermodynamically closed system.
As noted by Pfeifer, et al., \cite{BIPfeif} arbitrary partitions of the
total energy--momentum tensor are not particularly useful because 
only the total energy--momentum tensor is physically relevant.
\par
\section{Concluding Summary}
\par
Maxwellian continuum electrodynamics allows algebraic transformations 
that appear to be innocuous but change the conservation properties, the
relativistic properties, and the spacetime embedding of the coupled
equations of motion.
Identifying an arbitrarily large region of space in which the velocity
of light is $c/n$, instead of $c$, we defined a non-Minkowski material 
spacetime $(\bar x_0=ct/n,x,y,z)$.
Applying field theory in the material spacetime, we derived equations of
motion for the macroscopic electric and magnetic fields.
Requiring that we maintain a time-like coordinate $\bar x_0=ct/n$
prevents improper transformations of coordinates.
\par

\end{document}